\documentclass[11pt,twoside]{article}

\usepackage{asp2006}
\usepackage{epsf}
\usepackage{lscape}
\usepackage{graphicx}

\begin{document}
\title{The nature of ionized gas in early-type galaxies}   
\author{F. Annibali\altaffilmark{1}, A. Bressan\altaffilmark{1}, R. Rampazzo\altaffilmark{1}, W. W. Zeilinger\altaffilmark{2}}
\altaffiltext{1}{INAF - Osservatorio Astronomico di Padova, Vicolo dell'Osservatorio 5, I - 35122 Padova-ITALY }
\altaffiltext{2}{Institut f\"{u}r Astronomie, Universit\"{a}t Wien, T\"{u}rkenschanzstr. 17, A-1180 Wien, Austria}

\begin{abstract} 

We present a study of the ionized gas in a sample 
of 65 nearby early-type galaxies, for which we have acquired
optical intermediate-resolution spectra.
Emission lines are detected in $\sim$ 89\% of the sample.
The incidence of emission appears independent from the E or S0 morphological classes.
According to classical diagnostic diagrams, 
the majority of the galaxies are LINERs.
However, the galaxies tend to move toward the ``Composites''
region (at lower  [NII]$\lambda$6584/H$\alpha$ values)
as the emission lines are measured at larger 
galacto-centric distances. This suggests that different ionization 
mechanisms may be at work in LINERs.

\end{abstract}



\section{Introduction}

Early-type galaxies (ETGs) have long been considered to be inert stellar systems,
essentially devoid of gas and dust. However, this view has radically changed 
since a number of imaging and spectroscopy studies from both  
the ground and space have revealed the presence of a multiphase 
interstellar medium (ISM): 
a hot ($T \sim 10^6 - 10^7$ K), X-ray emitting halo \citep{fab92}; 
a warm ($T \sim 10^4$ K)  component (often referred to as ``ionized gas'')  
\citep{ph86}; and even cooler components detected in the MIR 
\citep{bre06} and in HI and CO  \citep{sad02}.
Ionized gas is detected in 40-80 \% of early-type galaxies via its
 optical emission lines \citep{sarz06,yan06,ser08}. 
 
Despite the number of studies, several issues remain still open. The 
first question is the origin of the ISM in ETGs. 
Evidence for external acquisition comes from narrow-band imaging studies,
often showing gas/star misalignment  \citep{bus93},
and from  kinematical studies, often showing gas/star
angular momentum decoupling \citep{caon00}.

The second still open issue concerns the ionizing source of the 
warm gas. Optical spectroscopic studies show that ETGs 
are typically classified as Low-Ionization Emission-line Regions (LINERs) 
according to their emission line ratios \citep{ph86}.
However, there is still strong debate about the ionization mechanism in LINERs.
Actually, the most viable excitation mechanisms are: 
low accretion-rate AGN \citep{kew06}, photoionization by old post-asymptotic
giant branch (post-AGB) stars \citep{bin94}, fast shocks \citep{ds95}.

\section{The Sample}

With the aim of understanding the nature of ionized gas in ETGs,
we acquired intermediate-resolution (FWHM $\approx$7.6~\AA\ at 5550~\AA) 
optical spectra (3700 - 7250 \AA)  for a sample of 65 nearby E/S0 galaxies. 
For a detailed description of the sample we refer to \cite{ramp05} (Paper~I) 
and \cite{ann06} (Paper~II).
For the derivation of ages, metallicities, and [$\alpha$/Fe] 
ratios, we refer to \cite{ann07} (Paper~III).
Here we recall that the sample was selected from a compilation of galaxies showing 
ISM traces in at least one of the following bands:
IRAS 100 $\mu$m, X-ray, radio, HI and CO \citep{rob91}.
Because of the selection criteria, the sample is biased toward the presence 
of emission lines. The galaxies are mainly located in low
density environments.
In the present study, we will use the spectra extracted in 4 annuli of increasing 
galacto-centric distance 
(0 $\leq$ r $\leq$r$_e$/16, r$_{e}$/16 $\leq$ r $\leq$r$_e$/8, r$_{e}$/8
$\leq$ r $\leq$r$_e$/4, and r$_{e}$/4 $\leq$ r $\leq$r$_e$/2).

\begin{figure}
\plotone{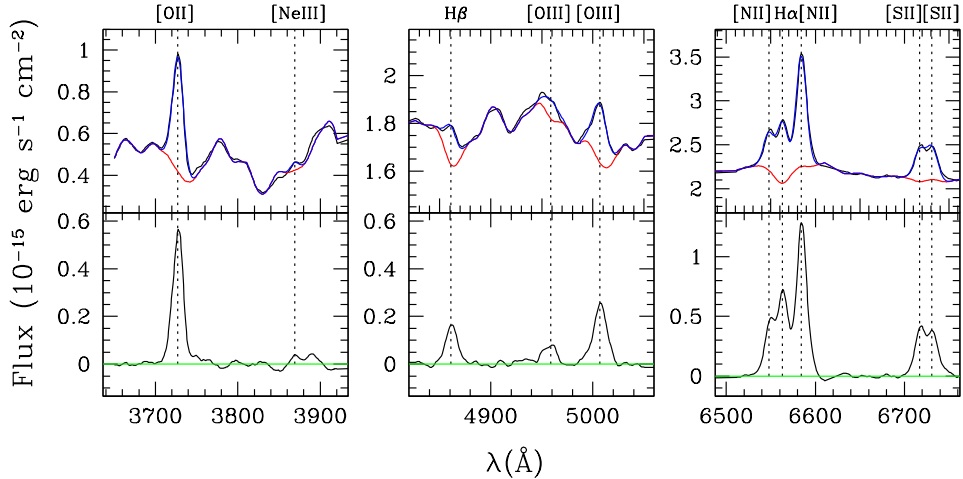}
\caption{Spectrum of IC~1459 at ${\rm r<r_e/16}$. Top panel: 
galaxy spectrum, normalized SSP model, and total fit obtained by adding emission lines 
to the underlying stellar continuum. 
Bottom panel: residual spectrum.}
 \label{fig1}
\end{figure}

\section{Emission lines}

The emission line fluxes were measured on residual spectra 
obtained by subtracting the stellar population contribution to the 
observed galaxy spectra. To this purpose, 
we used new SSPs (Bressan, unpublished; \cite{cha09}).
The SSPs were computed from the isochrones of \cite{bert94}, 
with the revision of \cite{bre98} including a new AGB mass-loss treatment.
In the optical domain, the SSPs are based on the MILES spectral library \citep{sb06}. 
The galaxy continua were fitted with the new SSPs, 
smoothed to match the instrumental resolution and the galaxy velocity dispersion,
through a $\chi^2$ criterion. In the fit, we considered selected spectral regions 
chosen to include both features particularly sensitive to age
(the relative strength of the Ca II H + K lines,  the 4000 \AA \ break, the H$\gamma$ and H$\delta$ lines), and features more sensitive  to metallicity (the Fe lines at $\lambda$ 4383, 4531, 5270, 5335, and the Mg absorption features around $\lambda$ 5175).
The residual spectrum around each line was derived 
by normalizing the SSP in two continuum bands adjacent to
the line of interest.
The line fluxes were determined by 
fitting the residual spectrum with Gaussian curves 
of variable width and intensity.
 As an example, we show the stellar continuum subtraction 
and the line fit for IC~1459 in Fig.~\ref{fig1}.
The emission lines were corrected for extinction through the observed ${\rm F(H\alpha) / F(H\beta)}$ flux ratio, assuming an  intrinsic value of  $\approx$ 3.1 valid for AGN-like objects (Osterbrock 1989).
Some galaxies in the sample present very large reddening, up to ${\rm E(B-V)  \sim 1.5}$ or even more.
Since the observed continuum is incompatible with such large values, the 
dust must be patchy.

Emission lines are detected in 58 out of 65 galaxies ($\sim$ 89 \% of the sample).
The incidence of emission appears independent from the E and S0 morphological classes.
The nebular emission is stronger in the galaxy center, and decreases moving toward the 
more external annuli.

\section{Diagnostic Diagrams}

\begin{figure}[ht!]
\vspace{-1.cm}
\plotone{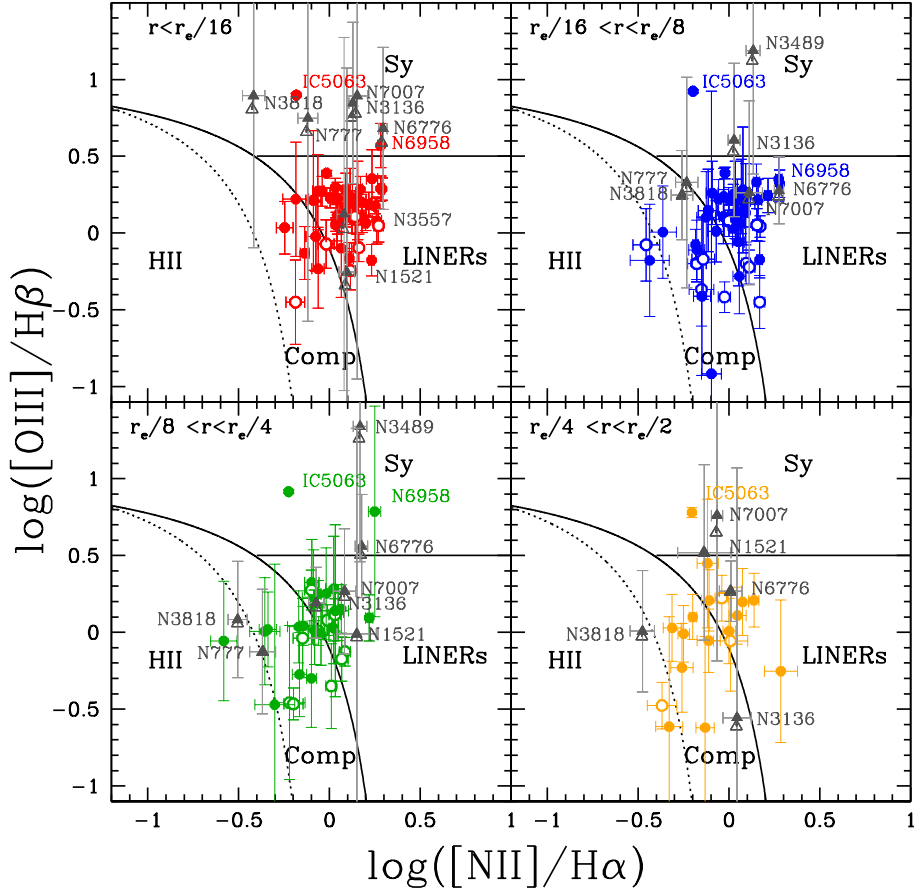}
\caption{Extinction corrected diagnostic diagram 
for our sample in 4 annuli of increasing galacto-centric distance.
 Empty circles denote the galaxies with problematic reddening derivation. 
 Triangles are for galaxies with ${\rm E(B-V)>1.5}$. 
 Full and empty triangles indicate observed and reddening-corrected
 line ratios, respectively.
 The solid curve is the ``maximum starburst line'' of \cite{kew01}, while
 the dashed line indicates the empirical division between pure star-forming galaxies and
 ``Composite'' objects. The horizontal line at 0.5  separates Seyfert and LINER galaxies.}
 \label{fig2}
\end{figure}

Galaxies were classified through the standard [OIII]$\lambda$5007/ H$\beta$ versus 
[NII]$\lambda$6583/ H$\alpha$ diagnostic diagram, first proposed by \cite{bpt} (hereafter BPT). 
The BPT diagram for our galaxy sample, in the 4 annuli, is shown in Fig.~\ref{fig2}.
The majority of galaxies are classified as LINERs.
Only IC~5063 is classified as a {\it bona fide} Seyfert galaxy.
A few galaxies, namely NGC~3489, NGC~3818, NGC~3136, 
NGC~6776, NGC~7007, NGC~777, and NGC~6958, fall in the Seyfert region, but are consistent 
with a LINER classification within the errors. 
Some galaxies (NGC~3258, NGC~4552, NGC~5193, NGC~5328, NGC~6721, NGC~6876, IC~2006) fall in the ``Composites'' region (see Fig.~\ref{fig2}), and possibly contain a combined contribution
from both star formation and AGN. 
We notice that, from the center outwards, the bulk of the galaxies 
move left-down in the BPT diagram. An increasing number of galaxies pass from 
LINERs to ``Composite'' objects as the emission lines are measured at larger 
galacto-centric distances.
This suggests that different ionization mechanisms
may be at work in LINERs.

\end{document}